\documentclass[sigconf]{acmart}

\usepackage{amsthm}
\usepackage{amsmath}
\usepackage{tabularx}  
\usepackage{graphicx}
\usepackage{subcaption}
\usepackage{hyperref}
\usepackage{multirow}
\usepackage{enumitem}
\usepackage{hyperref}

\usepackage{booktabs}
\newtheorem{definition}{Definition}

\copyrightyear{2025}
\acmYear{2025}
\setcopyright{cc}
\setcctype{by}
\acmConference[CIKM '25]{Proceedings of the 34th ACM International Conference on Information and Knowledge Management}{November 10--14, 2025}{Seoul, Republic of Korea}
\acmBooktitle{Proceedings of the 34th ACM International Conference on Information and Knowledge Management (CIKM '25), November 10--14, 2025, Seoul, Republic of Korea}\acmDOI{10.1145/3746252.3761263}
\acmISBN{979-8-4007-2040-6/2025/11}

\begin{document}

\title%
{Evaluating and Addressing Fairness Across User Groups in Negative Sampling for Recommender Systems}

\author{Yueqing Xuan}
\affiliation{
  \institution{RMIT University}
  \city{Melbourne}
  \state{Victoria}
  \country{Australia}
}
\email{yueqing.xuan@student.rmit.edu.au}
\orcid{0000-0002-9365-8949}

\author{Kacper Sokol}
\affiliation{
  \institution{ETH Zurich}
  \city{Zurich}
  \country{Switzerland}
}
\email{kacper.sokol@inf.ethz.ch}
\orcid{0000-0002-9869-5896}

\author{Mark Sanderson}
\affiliation{
  \institution{RMIT University}
  \city{Melbourne}
  \state{Victoria}
  \country{Australia}
}
\email{mark.sanderson@rmit.edu.au}
\orcid{0000-0003-0487-9609}

\author{Jeffrey Chan}
\affiliation{
  \institution{RMIT University}
  \city{Melbourne}
  \state{Victoria}
  \country{Australia}
}
\email{jeffrey.chan@rmit.edu.au}
\orcid{0000-0002-7865-072X}

\renewcommand{\shortauthors}{Yueqing Xuan et al.}

\begin{abstract}
Recommender systems trained on implicit feedback data rely on negative sampling to distinguish positive items from negative items for each user. %
Since the majority of positive interactions come from a small group of active users, negative samplers are often impacted by data imbalance, leading them to choose more informative negatives for prominent users while providing less useful ones for users who are not so active. %
This leads to inactive users being further marginalised in the training process, thus receiving inferior recommendations. %
In this paper, we conduct a comprehensive empirical study %
demonstrating that state-of-the-art negative sampling strategies provide more accurate recommendations for active users than for inactive users. %
We also find that increasing the number of negative samples for each positive item improves the average performance, but the benefit %
is distributed unequally across user groups, with active users experiencing performance gain while inactive users suffering performance degradation. %
To address this, %
we propose a group-specific negative sampling strategy that assigns smaller negative ratios to inactive user groups and larger ratios to active groups. %
Experiments on eight negative samplers show that our %
approach %
improves user-side fairness and performance when compared to a uniform global ratio. %
\end{abstract}



\maketitle

\section{Introduction}

Recommender systems are widely used to help users discover items of interest~\cite{ricci2010introduction}, with Collaborative Filtering (CF) methods~\cite{su2009survey} being among the most prevalent. These recommenders learn user preferences from implicit feedback data, where the observed user--item interactions are treated as positive data. Negative sampling is commonly used to draw negative items from a large set of items that a user has never interacted with~\cite{rendle2012bpr}. 

State-of-the-art negative samplers, such as MixGCF~\cite{huang2021mixgcf} and DENS~\cite{lai2023disentangled}, select negative samples for each user by utilising their positive feedback data. %
However, data imbalance remains a significant challenge for recommenders~\cite{li2021user}. For example, in the classic ML~1M dataset, over 60\% of interactions comes from only 25\% of users. This concentration of data leads to negative samplers selecting more informative negative items for active users with abundant positive interactions.
On the other hand, the remaining 75\% of users is less active and often lacks sufficient data for the samplers to identify informative negatives, which exacerbates the recommendation disparities for these under-represented users. %

In most recommendation datasets, a significant percentage of users can be considered inactive, where we define inactivity as having limited interactions with the system rather than being entirely new with no prior activity. 
A non-satisfactory service for such users may discourage them from engaging further with the platform. %
Moreover, inactive users often include a large demographic of individuals who have limited digital literacy or access to digital services, e.g., the elderly or those from disadvantaged socio-economic backgrounds~\cite{thomas2023measuring}. Improving their experience is crucial for fair recommendation and inclusive digital transformation~\cite{thomas2023measuring}. Consequently, assessing performance disparity across user groups is essential to ensuring that the experience of disadvantaged or under-represented users with limited engagement is not overshadowed by the preferences of a small group of active users.

Current work on evaluating user-side fairness in recommendation found that active users typically receive more accurate results than inactive users~\cite{li2021user,fu2020fairness}. However, these studies are limited to %
the random sampling strategy. %
Given the widespread adoption of user-level, personalised negative samplers, it is important not only to optimise their effectiveness but also to ensure that all users, regardless of their activity level, are provided with equally informative negative samples. %
To the best of our knowledge, the role of negative samplers in performance disparity among user groups with varying activity levels is yet to be addressed. %

Literature also shows that raising the number of negative items sampled for each positive interaction %
consistently enhances the average performance of CF models~\cite{huang2021mixgcf,chen2023revisiting}. Again, this result %
has only been validated for the random sampling strategy. It remains to be seen whether all negative samplers and all user groups benefit from a higher negative sampling ratio. Moreover, studies on negative sampling in contrastive learning find that a bigger negative ratio can introduce excessive false negatives and harm the model training, whereas a smaller ratio may not provide sufficient information for the model to learn an accurate decision boundary~\cite{menon2020can,chen2023revisiting}. %
However, this line of work applies a uniform sampling ratio across all users, overlooking the existence of distinct sub-populations with varying activity levels. %
It remains unclear whether a dataset-wide negative ratio leads to optimal performance for all groups, or if different groups achieve their best performance at different ratios.

In this paper, we present a comprehensive empirical study to uncover and address performance disparities in recommender systems arising from negative sampling. We evaluate a range of representative sampling strategies across user groups with varying activity levels, and reveal how existing approaches inadvertently amplify unfairness in recommendation quality. %
We structure our investigation around four research questions (RQ).
\textbf{RQ1} examines whether different negative samplers deliver equal recommendation quality to different user groups %
for a fixed recommender.
\textbf{RQ2} explores whether increasing negative sampling ratios enhances or harms the dataset-level performance of these samplers. %
\textbf{RQ3} investigates whether such dataset-wide performance gain (or loss) is equally distributed across user sub-populations, or whether inactive users are at a disadvantage. %
Our results uncover an under-explored fairness issue: although increasing negative sampling ratios generally improves dataset-wide performance, it systematically disadvantages inactive users while masking the unfairness behind aggregate gains. 

Building on these findings, \textbf{RQ4} further investigates whether assigning group-specific negative sampling ratios, instead of using a global setting, improves performance, particularly for inactive users. To address RQ4, we propose a hyperparameter optimisation strategy that identifies optimal negative sampling ratios tailored to each user group. Extensive experiments show that using group-specific ratios not only improves overall recommendation accuracy but also substantially enhances performance for inactive users.

The contributions of this paper are three-fold. %
\begin{enumerate}[leftmargin=*,topsep=0pt]%
    \item We provide the first in-depth, empirical analysis of eight widely used negative sampling strategies on a user-group level, revealing that none of them achieves equitable performance across user groups. This performance bias, rooted in data imbalance, persists across recommenders and datasets. 
    \item We demonstrate that increasing negative sampling ratios can improve dataset-wide performance of %
    samplers, but the performance gain disproportionately benefits active users,
    with inactive users experiencing performance degradation. This insight calls for more equitable negative sampling strategies.  %
    \item %
    We are the first to identify that different user groups require different optimal negative sampling ratios -- a smaller ratio suffices for inactive users, while a higher ratio is needed to support active users. We operationalise this finding through a hyperparameter optimisation framework to assign tailored ratios to different user groups. 
    Extensive experiments confirm that using group-specific ratios outperforms the global ratio across all samplers, improving both recommendation accuracy and user-side fairness. %
\end{enumerate}

\section{Related Work and Preliminaries}

\paragraph{Negative Sampling for Recommendation}
Recommenders are often trained on users' implicit feedback data, where observed interactions indicate positive preferences, and unobserved data contain both unlabelled potential positives and true negatives.
Negative sampling strategies %
are used to draw negative items from the unobserved data. %
\emph{Static} negative sampling approaches utilise a fixed sampling distribution for each user. The simplest approach is to sample negative instances from non-interacted items uniformly at random~\cite{rendle2012bpr}. Otherwise, MCNS~\cite{yang2020understanding} uses a negative sampling distribution that approximates the positive distribution, and PNS~\cite{chen2017sampling} is an item popularity-based sampling method to mitigate popularity bias and help with long-tail recommendation. %

\emph{Hard} negative sampling approaches select %
negative items that have high probability of being misclassified as positives by a recommender for each user~\cite{zhang2013optimizing,rendle2014improving,wang2017irgan}. These negative instances are close to decision boundaries thus allow more precise delineation of user preferences. 
DNS~\cite{zhang2013optimizing} dynamically selects negative items with the highest prediction scores, and SRNS~\cite{ding2020simplify} samples items with high prediction scores and high variance to avoid false negative instances. Additionally, MixGCF~\cite{huang2021mixgcf} synthesises hard negatives by mixing information from positive instances into negative samples for graph-based recommenders. %
DENS~\cite{lai2023disentangled} disentangles relevant and irrelevant factors of each item to reveal a user's preference and then selects the negative samples by contrasting their relevant factors with those of positive items. Notably, all these methods aim to maximise the overall utility but ignore the performance variation between different user groups. Our work is the first attempt to identify recommendation performance disparity across user groups in the context of negative sampling.     %

\paragraph{Fairness in Recommender Systems}
Fairness research in recommender systems %
addresses either the user~\cite{leonhardt2018user,wang2021user,anelli2023challenging} or the item~\cite{abdollahpouri2017controlling,zhu2021fairness} side, at the same time striving to %
keep the accuracy high~\cite{rahmani2022unfairness}. %
On the \emph{user side}, inactive users have been shown to receive lower quality recommendations 
due to their insufficient training data~\cite{fu2020fairness}, leading to a fairness-constrained approach via re-ranking for recommenders based on knowledge graph. 
\citet{anelli2023challenging} observed that active users receive more accurate recommendations from graph-based and classic CFs. 
However, counter-intuitively, users with many historical interactions suffer from relatively poor recommendations; instead, recommenders give better results to users with more recent interactions~\cite{ji2022loyal}. 
To achieve user-side fairness, existing work often utilises in-processing methods, such as model training regularisation or adversarial learning~\cite{li2021user,jin2023survey,pitoura2022fairness}. %
Pre-processing strategies to address performance disparity, e.g., using data (re)sampling, remain under-explored~\cite{jin2023survey}.

In terms of \emph{item-side} fairness, negative sampling has been used to address item popularity bias. 
\citet{wu2024effectiveness} proposed the sampled \textit{softmax} loss function, %
which benefits long-tail recommendation. %
\citet{chen2023fairly} found that items from the majority class are more likely to be selected as negative samples; they addressed this problem by adaptively adjusting the negative sampling distribution for each item group. However, the influence of negative samplers on the performance received by different user groups remains unexplored. Our work addresses this gap and focuses on user-side fairness in the context of negative sampling. %

\paragraph{Preliminaries}

Our work focuses on recommendation using implicit feedback data, 
where each data sample is indexed by a user ID, an item ID and contains a binary label indicating whether a user has interacted with an item or not. 
Let $U$ and $I$ respectively denote the set of users and items in a dataset, %
and $Y =\{(u,i) \, | \, u\in U, \, i\in I \}$ the set of observed interactions where each pair is a positive interaction between a user $u$ and an item $i$. 
A scoring function $\hat{y}(u,i)\rightarrow \mathbb{R}^+$ predicts the preference of $u$ towards $i$. 
$I_u^+$ denotes the interacted (positive) item set for $u$, and $I_u^-=I \setminus I_u^+$ denotes $u$'s non-interacted item set. CF models are often optimised through 
the pairwise Bayesian Personalised Ranking (BPR) loss~\cite{rendle2012bpr}, where each positive interaction $(u,i)$ is paired with one negative item $j\in I_u^-$, namely: 
\[
    \mathcal{L}_{BPR} = - \sum_{(u,i)\in Y, \, j\in I_u^-}
    \ln \sigma \left(\hat{y}(u,i) -\hat{y}(u,j)\right)\text{,}
\]
where $\sigma$ is the sigmoid function. Negative sampling strategy is used to sample $j$ from $I_u^-$ for each $(u,i)$. 
Minimising $\mathcal{L}_{BPR}$ encourages the predicted score on a positive item to be higher than on its paired negative item.
When a positive item is paired with $\mathcal{K}$ negative items,  
the BPR loss with multiple negatives is formulated as: 
\begin{equation}\label{eq:bpr_p}
    \mathcal{L}_{BPR+} = -\sum_{(u,i)\in Y} \sum_{j\in I_u^-}^\mathcal{K} \ln \sigma(\hat{y}(u,i)-\hat{y}(u,j))\text{.}
\end{equation}

\section{Evaluating Negative Samplers}

In this section we introduce the first three research questions on evaluating performance disparities for different negative samplers. %
Before proceeding, we first present our user partitioning strategy. %

\paragraph{User Partitioning} %
To address group fairness, it is essential to partition users into groups upon specific attributes. In implicit recommendation settings, %
sensitive attributes such as gender are often unavailable~\cite{fu2020fairness}. Following established practice~\cite{fu2020fairness,anelli2023challenging}, %
we split users based on their number of positive interactions (i.e., activity level) recorded in the training set, %
with each group containing an equal number of users. We denote the total number of groups as $n$, with $U=\{\textsc{u}_1, \textsc{u}_2, \ldots, \textsc{u}_n\}$. 
Consistent with established terminology for user grouping based on activity level, we refer to them as relatively \emph{inactive} vs.\ \emph{active} users~\cite{fu2020fairness,anelli2023challenging}. We later confirm that each user group exhibits significantly different activity level, reflecting genuinely distinct user populations.
In what follows, we first present our findings using quartile-based user splits (i.e., $n=4$) and then show that our findings hold for other values of $n$. %

We choose user activity level as the partitioning criterion for four reasons. %
One, user activity level is a directly observable property and is widely used %
in fair recommendation research as a proxy for sensitive attributes~\cite{fu2020fairness,li2021user,ji2022loyal,anelli2023challenging,dong2023newer}. %
Two, it is also a representative attribute, as in implicit feedback data it reflects both the number of unique items a user has interacted with and the diversity of their interactions. %
Three, previous work suggests that user activity level can directly entail subpar treatment by recommenders~\cite{fu2020fairness}; later in this work we also confirm that it is strongly correlated with disparate performance in sampling-based recommenders.
Four, we further partition users by leveraging their temporal information %
in Section~\ref{sec:ablation}; nonetheless, these alternative criteria do not yield equally informative results, evidentiating that activity level is the most appropriate partitioning criterion. %

\paragraph{RQ1: Performance Disparity}%

After forming user groups, we first aim to assess whether different negative samplers, when paired with the same backbone recommender, provide equal recommendation quality across user groups. %
While existing studies evaluate performance disparities among recommenders under the random negative sampling strategy, they overlook the impact of different negative samplers on fairness under a fixed recommender~\cite{li2021user,ji2022loyal}.
This oversight is crucial because hard negative samplers often personalise the negative item distribution for each user based on their positive interaction data, and given that some users contribute disproportionately to the interaction data, these samplers may amplify data imbalance issues.
In particular, users with limited positive data might receive more false negatives (i.e., less informative negative samples), adversely impacting their recommendation quality. 
Given that hard samplers can increase a recommender's overall performance compared to static samplers~\cite{chen2023revisiting}, we investigate whether this performance improvement disproportionately benefits active users, thereby exacerbating performance disparities between active and inactive user groups.

\paragraph{RQ2: Performance Change for Larger Ratios}%

Existing negative samplers typically select one negative item per positive interaction~\cite{rendle2012bpr,he2017neural,chen2023revisiting}. However, the impact of negative sampling ratio $\mathcal{K}$ on the average utility of a sampler has not been thoroughly examined. 
To address this, we explore how the performance of different negative samplers, integrated with the same recommender, changes as the negative sampling ratio increases. %
Some work argued that increasing $\mathcal{K}$ improves the performance as the recommender learns to distinguish among multiple negatives simultaneously~\cite{huang2021mixgcf,chen2023revisiting}; %
however, their analyses are limited to a specific sampler (MixGCF) or the random sampler. %
On the other hand, \citet{wu2021rethinking} suggested that %
a medium value of $\mathcal{K}$ is optimal because small $\mathcal{K}$ underutilises useful information 
while large $\mathcal{K}$ introduces label noise that harms model training. %
Yet their findings are based solely on random sampling with a single recommender.
It remains unclear whether all negative samplers benefit from an increased ratio. %
Additionally, it is unknown if increasing $\mathcal{K}$ alters the relative performance ranking of different samplers. %
For example, a sampler that outperforms another at $\mathcal{K}=1$ may not maintain its advantage at $\mathcal{K}=8$ or higher. This raises a critical question about the consistency of sampler rankings, particularly since current studies benchmark samplers without varying $\mathcal{K}$. %

\paragraph{RQ3: Performance Change Distribution}%

After studying how aggregate performance of negative samplers (measured by averaging utility metrics across all users) changes when $\mathcal{K}$ increases (RQ2), we note that this evaluation approach ignores how the performance change is distributed across user sub-populations. %
In the context of fairness, an important notion is demographic parity, which ensures that disadvantaged and advantaged users have equal chances of success~\cite{zemel2013learning}.
Inspired by this concept, we explore whether different user groups equally experience the performance gain (or loss) when the average performance of negative samplers improves (or deteriorates) after increasing $\mathcal{K}$. %
We formalise this property in Definition~\ref{def:change-disp}. %

\begin{definition}[Performance Change Distribution Disparity]\label{def:change-disp}
A negative sampler fairly distributes performance gain (or loss) when different user groups experience the same degree of performance change as negative sampling ratio $\mathcal{K}$ increases. Namely, 
\begin{equation*}
\begin{aligned}
    & \Delta(\textsc{u}) = \frac{M(\textsc{u})|\mathcal{K}_2 - M(\textsc{u})|\mathcal{K}_1}{M(\textsc{u})|\mathcal{K}_1} \quad && \text{for}\;\; \mathcal{K}_2 > \mathcal{K}_1 \\
    & \Delta(\textsc{u}_1) = \Delta(\textsc{u}_2) \qquad && \forall \quad \textsc{u}_1,\textsc{u}_2 \in U\text{,}   
\end{aligned}
\end{equation*}
where M is a utility metric and $M(\textsc{u})|\mathcal{K}$ is the performance for group \textsc{u} under ratio $\mathcal{K}$.
\end{definition}

When the performance gain is equally distributed, each user group benefits comparably, and performance disparity remains unchanged, i.e., it does not worsen. %
Nonetheless, it may be desirable for the inactive group -- which we later find to be at a disadvantage under $\mathcal{K}=1$ (RQ1) -- to receive a greater degree of gain than the active group under a higher $\mathcal{K}$, helping to reduce the performance disparity. In the worst case, the advantaged (active) group continues to enjoy improved performance, while the disadvantaged (inactive) group suffers performance degradation, exacerbating the disparity. 
To promote fairness, it is thus crucial to examine how the relative performance changes for individual user groups as $\mathcal{K}$ increases.

Definition~\ref{def:change-disp} can be used as a pairwise metric to compare two user groups under two different $\mathcal{K}$ values. When there are more than two groups (e.g., $n=4$ in this work) and multiple $\mathcal{K}$ values (e.g., $\mathcal{K} \in \{1, 2, 3, \ldots\}$), we further apply Kendall's $W$ (coefficient of concordance~\cite{gibbons2014nonparametric}) to assess whether the most active user group consistently receives the greatest relative performance gain, while less active groups receive smaller relative gains. Specifically, Kendall's $W$ evaluates the characteristic outlined in Definition~\ref{def:ranking}. %

\begin{definition}[Unequal Distribution Persistence]\label{def:ranking}
Assuming that the most active user group receives the highest degree of performance gain as $\mathcal{K}$ increases, followed by the second most active group with the second highest degree of gain, and so on, Kendall's $W$ measures how consistently this rank order is maintained as we gradually increase the value of $\mathcal{K}$. %
\end{definition}

Under Definition~\ref{def:ranking}, when $W=1$, the performance gain is consistently distributed such that the most active user group receives the largest relative gain, followed by the second most active group with the second largest gain, and so on, with the least active group receiving the smallest relative gain. %
In contrast, when $W=0$, the ranking of performance gain is entirely random, i.e., no user group consistently receives the highest or lowest relative gain.
We are thus interested in how close $W$ is to 1 %
as this indicates a persistent unequal distribution of performance gains, which represents the worst-case scenario in terms of fairness.

\section{Experimental Setup}

\paragraph{Negative Samplers}
We select eight popular negative sampling strategies; the first two are \emph{static} and the remaining six are \emph{hard}. %
\begin{description}[topsep=0pt]%
    \item [Random Negative Sampling (RNS)]
    follows the uniform distribution to randomly sample negative items~\cite{rendle2012bpr}. 
    \item [Popularity-based NS (PNS)]
    assigns higher probability to popular unobserved items~\cite{chen2017sampling}.
    \item [Dynamic NS (DNS)]
    selects items with the highest prediction score from a candidate set of negative items~\cite{zhang2013optimizing}.
    \item [Disentangled NS (DENS)]
    disentangles relevant and irrelevant factors of items and identifies the best negative samples by contrasting the relevant factors~\cite{lai2023disentangled}. %
    \item [Adaptive Hardness NS (AHNS)]
    selects negative items with different hardness levels adaptively for each positive item~\cite{lai2024adaptive}. %
    \item [DNS(M,~N)]
    extends DNS by using $M$ to control sampling hardness and $N$ to determine the size of the candidate set~\cite{shi2023theories}.
    \item [Importance Resampling (AdaSIR)]
    reuses historical hard samples with importance resampling to maintain the candidate set~\cite{chen2022learning}.
    \item [MixGCF]
    integrates multiple negatives to synthesise a hard negative through positive mixing and hop mixing~\cite{huang2021mixgcf}.
\end{description}%
We do not experiment with methods like SRNS~\cite{ding2020simplify} or MCNS~\cite{yang2020understanding} as they are widely shown to underperform~\cite{huang2021mixgcf,chen2023revisiting,lai2023disentangled,lai2024adaptive}. 
We leave experiments on adversarial-based samplers~\cite{wang2017irgan,park2019adversarial} for future work. %

Among the eight samplers, MixGCF works only with graph-based recommenders; the remaining methods are model-agnostic. %
We pair each sampler with two representative sampling-based recommenders: Matrix Factorisation (MF) and Light Graph Convolution Network (LightGCN)~\cite{he2020lightgcn} (except for MixGCF on MF). This setup ensures that our findings generalise across recommenders. %

\begin{table}[t]
    \centering 
    \footnotesize
    \setlength{\tabcolsep}{3pt}%
    \caption{Dataset statistics.}
    \begin{tabular}{@{}lrrrrl@{}}  
        \toprule
        Dataset & $|U|$ & $|I|$ & $|Y|$ & Density & Characteristics\\ \midrule
        Amazon Beauty \cite{he2016ups} & 22,363 & 12,101 & 198,502 & 0.07\% & small \& sparse  \\
        MovieLens (ML) 1M \cite{harper2015movielens} & 6,040 & 3,706 & 1,000,209  & 4.47\% & medium \& dense \\
        Book Crossing \cite{ziegler2005improving} & 15,798  & 38,093 & 585,579 & 0.10\% & medium \& sparse \\ 
        Yelp~2022 \cite{yelp_open_dataset}   & 287,116  & 148,523 & 4,392,169 &  0.01\% & large \& sparse \\
        \bottomrule
        \end{tabular}
    \label{tab:dataset}
\end{table}

\begin{figure*}[t]
    \centering
    \begin{subfigure}{0.22\linewidth}
        \includegraphics[height=.85\linewidth]{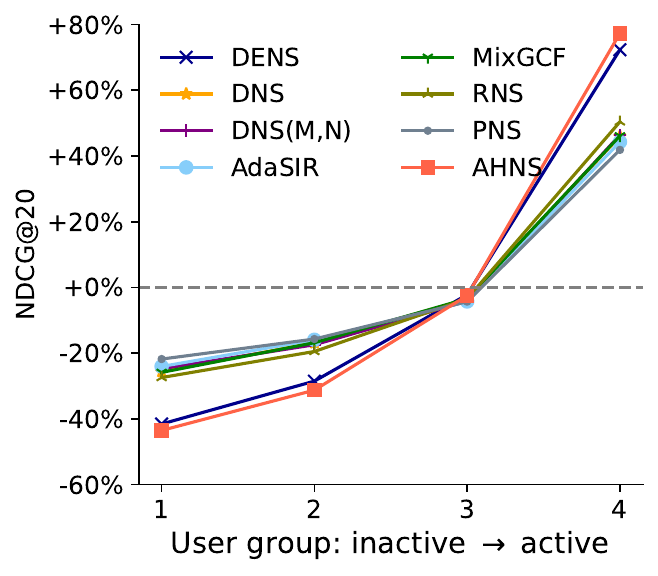}
        \caption{ML~1M.}
        \label{fig:ndcg_a}
    \end{subfigure}
\hspace{0.3em}
    \begin{subfigure}{0.22\linewidth}
        \includegraphics[height=.85\linewidth]{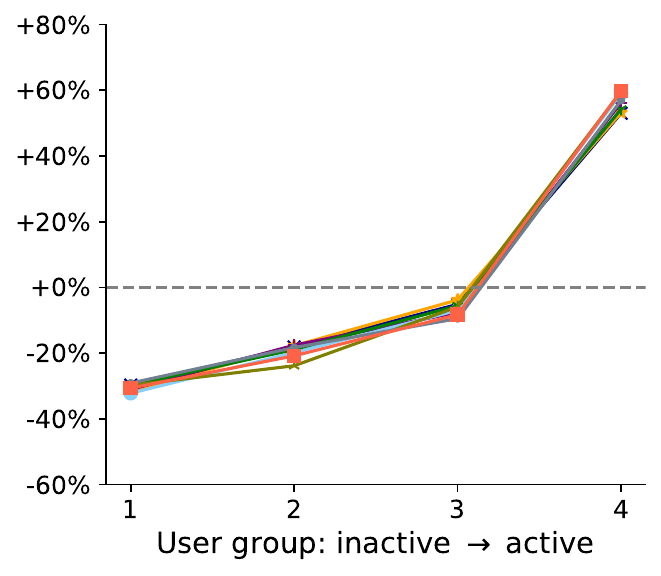}
        \caption{Amazon Beauty.}
        \label{fig:ndcg_b}
    \end{subfigure}
\hspace{0.3em}
    \begin{subfigure}{0.22\linewidth}
        \includegraphics[height=.85\linewidth]{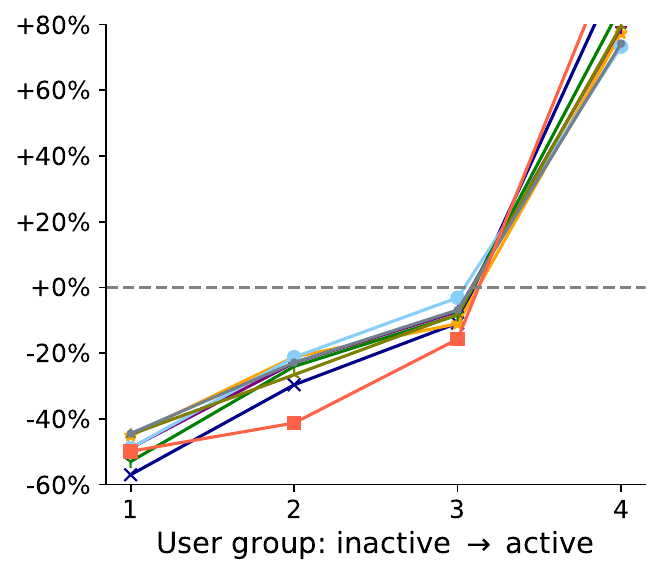}
        \caption{Book Crossing.}
        \label{fig:ndcg_c}
    \end{subfigure}
\hspace{0.3em}
    \begin{subfigure}{0.22\linewidth}
        \includegraphics[height=.85\linewidth]{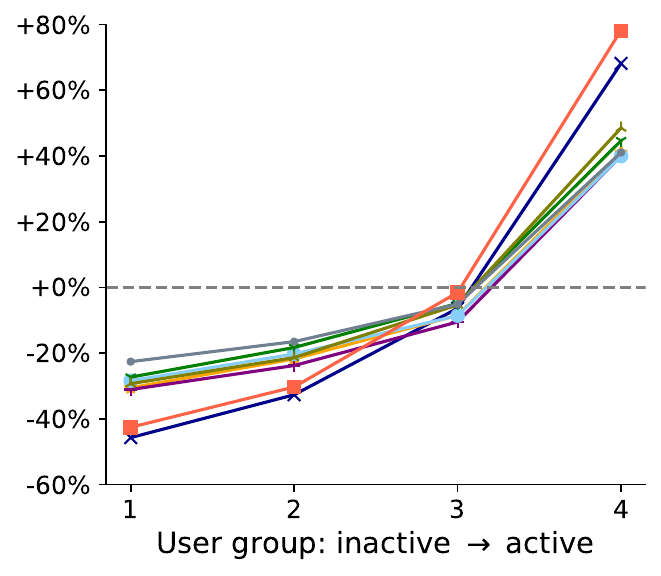}
        \caption{Yelp~2022.}
        \label{fig:ndcg_d}
    \end{subfigure}
\caption{Percentage variation between the NDCG of four user groups compared to the average NDCG value across all users (dashed line) for each dataset. All negative samplers are run with the LightGCN model.
    }
    \label{fig:one_k_user}
\end{figure*}

\paragraph{Datasets}
We use four representative public datasets with their details summarised in Table~\ref{tab:dataset}. Our dataset selection is informed by a prior survey~\cite{chin2022datasets}, ensuring a mix of small to large as well as sparse and dense datasets. 
We apply $5$-core filtering on Amazon Beauty and Yelp~2022 to exclude users and items with less than five interactions. Following common practice to create implicit feedback data, the item ratings are binarised to indicate whether a user has interacted with an item or not. Since all methods in our study are non-temporal,
we adopt a fixed-ratio data split, dividing each dataset into training, validation and test sets using a 70\%--10\%--20\% ratio~\cite{wang2019neural,yang2020understanding,huang2021mixgcf}.

\paragraph{Evaluation Metrics}
We evaluate the recommendation performance using two metrics: NDCG@$k$ and Recall@$k$. Following common practice, we set $k=20$~\cite{huang2021mixgcf,chen2023revisiting,chen2023fairly,lai2023disentangled,lai2024adaptive}. During evaluation, a recommender generates a ranked list for each user by ordering all items except those already in the user's training set, hence we do not use sampling during evaluation in line with best practice~\cite{krichene2020sampled}. %

\paragraph{Hyperparameter Setting}
We implement all our baseline models using PyTorch with the Adam optimiser. For all experiments, we fix the embedding size at 64 and batch size at 2,048. We conduct a grid search to find the optimal settings for each method. Specifically, the learning rate is searched over $\{0.0001,0.0005,0.001\}$, and the $L_2$ regularisation weight is tuned over $\{10^{-5}, 10^{-4}, 10^{-3}\}$. The candidate size $M$ used in hard negative samplers is searched in $\{4,6,8,10\}$. The number of layers in LightGCN and MixGCF is set to $3$. The details of additional hyperparameters used in each negative sampler are provided alongside our implementation in the dedicated code repository\footnote{%
\url{https://github.com/xuanxuanxuan-git/fair_ns}}.

\section{Empirical Findings for RQ1, RQ2 and RQ3}\label{sec:findings}
In this section, we address our first three research questions and summarise key findings. %
We also %
present ablation study results.

\subsection{Active Users Receive Better Recommendations (RQ1)}%

To answer RQ1, we compare the performance of negative samplers across different user groups.
Figure~\ref{fig:one_k_user} shows the percentage variation in performance (measured by NDCG) across quartiles relative to the average metric value over the entire dataset. It demonstrates how recommendation quality fluctuates among user groups. Note that the plots offer no insight into the overall recommendation utility, which is reported later in Table~\ref{tab:results}. %

For all the datasets, none of the negative samplers demonstrates fair recommendation behaviour. Regardless of the negative sampling technique, the recommenders disproportionately favour the most active user group (Group~4). %
Specifically, users in Group~4 enjoy over 60\% improvement in NDCG, whereas Groups~1--3 systematically suffer below-average performance. Moreover, AHNS exhibits greater penalisation for Groups~1--3 on ML~1M and Yelp~2022 compared to other methods. %
We posit that, since AHNS is an instance-wise sampling method, it personalises the sampling more accurately for the most active users but is less effective for inactive users.
The results for all the negative samplers with the MF model across all the datasets follow the same patterns. Full experimental results are available in the aforementioned code repository. %

We also observe that the extent of performance disparity varies across the datasets. Book Crossing exhibits the highest disparity, while Amazon Beauty shows the lowest. %
To understand these differences, we examine each group's contribution to the training data. As shown in Table~\ref{tab:data-dist}, the degree of data imbalance differs substantially across the datasets.
In all cases, Group 4 contributes more than 50\% of the training data; the contribution is as high as 80\% in Book Crossing, whereas Group 1 contributes only 3\% and Group 2 only 5\%. The relative lack of data for Groups 1--3 implies that the corresponding user preferences are barely captured, instead being dominated by the preferences of active users. A highly imbalanced data contribution exacerbates the performance discrepancies, e.g., as seen in Book Crossing. Moreover, although user groups differ significantly in their average activity levels within each dataset, the variation in the difference across groups is most extreme in Book Crossing. This also explains the performance gaps.

In summary, we find that for imbalanced data recommenders are biased towards active users who contribute their majority. %
The severity of this bias is closely tied to the degree of data imbalance. 
In most cases, %
using personalised item distribution in hard negative samplers does not further penalise under-represented users as compared to static samplers. %
Nonetheless, none of the hard samplers addresses the performance disparity, with active users being consistently advantaged across all the sampling methods we examined.

\begin{table}[b]
\centering
\footnotesize
    \setlength{\tabcolsep}{3pt}%
\caption{Statistics of user groups: proportion of total interactions (\%) and average number of interactions (avg). User groups are sorted by their activity level in ascending order -- Group 1 with the least and Group 4 with the most interactions. Each group has an equal number of users. %
}
    \begin{tabular}{@{}lrrrrrrrr@{}}
    \toprule
      & \multicolumn{2}{c}{Group 1}& \multicolumn{2}{c}{Group 2} & \multicolumn{2}{c}{Group 3} & \multicolumn{2}{c}{Group 4} \\ \cmidrule(lr){2-3} \cmidrule(lr){4-5}\cmidrule(lr){6-7} \cmidrule(lr){8-9}
      & \% & avg & \% & avg & \% & avg & \% & avg \\\midrule
     Amazon Beauty & 12\% & 3.20  & 16\% & 5.68  & 21\% & 8.52 &  51\% & 18.11  \\
     ML~1M &  5\% & 33.12   & 10\% & 66.23  & 21\% & 139.10  & 64\% & 423.93 \\
     Book Crossing  & 3\% &4.45  &  5\% & 14.83  & 11\% & 28.17  & 80\% & 100.82\\
     Yelp~2022 &  7\% & 4.28  & 11\% & 6.73 & 17\% & 10.40 &  65\% &39.77  \\
     \bottomrule
    \end{tabular}
    \label{tab:data-dist}
\end{table}

\begin{figure*}[t]
    \centering
    \begin{subfigure}{0.22\linewidth}
        \includegraphics[height=.85\linewidth]{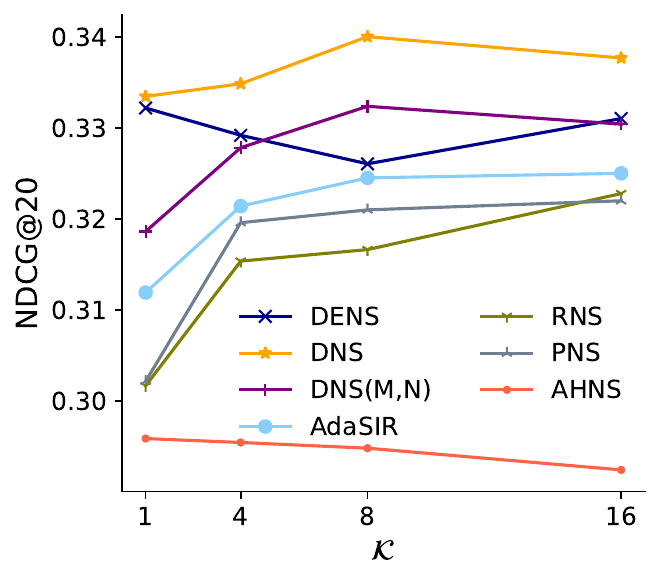}
        \caption{ML~1M -- MF.}
        \label{fig:ml_mf}
    \end{subfigure}
\hspace{0.4em}
    \begin{subfigure}{0.22\linewidth}
        \includegraphics[height=0.85\linewidth]{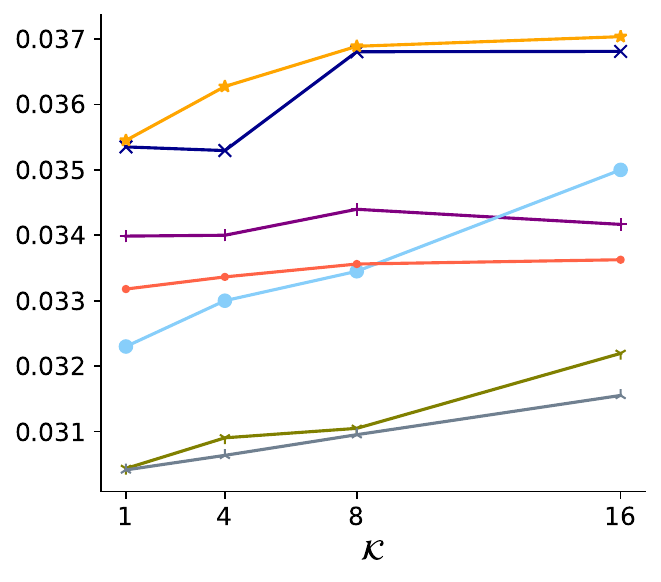}
        \caption{Amazon Beauty -- MF.}
        \label{fig:beauty_mf}
    \end{subfigure}
\hspace{0.4em}
    \begin{subfigure}{0.22\linewidth}
        \includegraphics[height=0.85\linewidth]{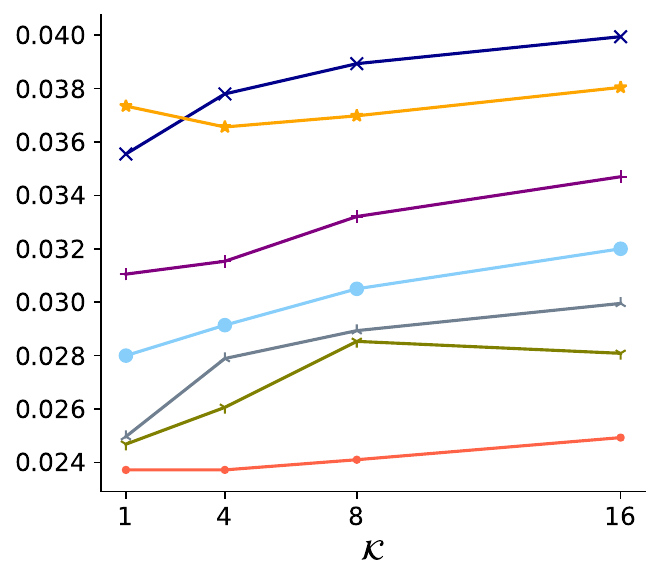}
        \caption{Book Crossing -- MF.}
        \label{fig:book_mf}
    \end{subfigure}
\hspace{0.4em}
    \begin{subfigure}{0.22\linewidth}
        \includegraphics[height=0.85\linewidth]{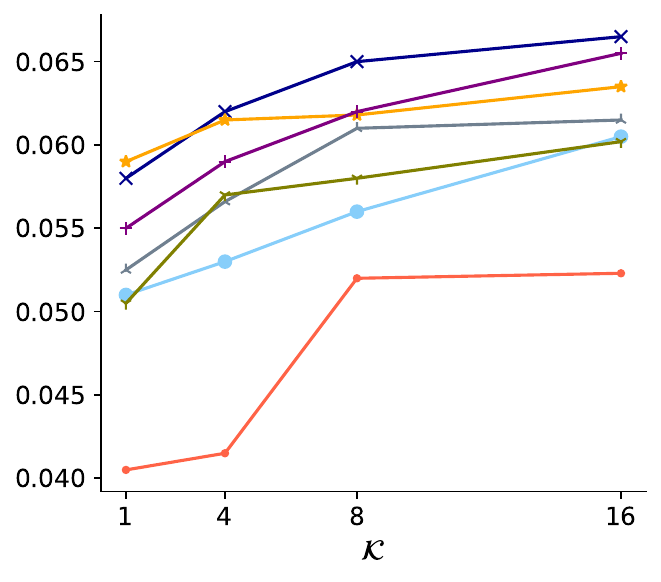}
        \caption{Yelp~2022 -- MF.}
        \label{fig:pinterest_mf}
    \end{subfigure}
\\
    \begin{subfigure}{0.22\linewidth}
        \includegraphics[height=0.85\linewidth]{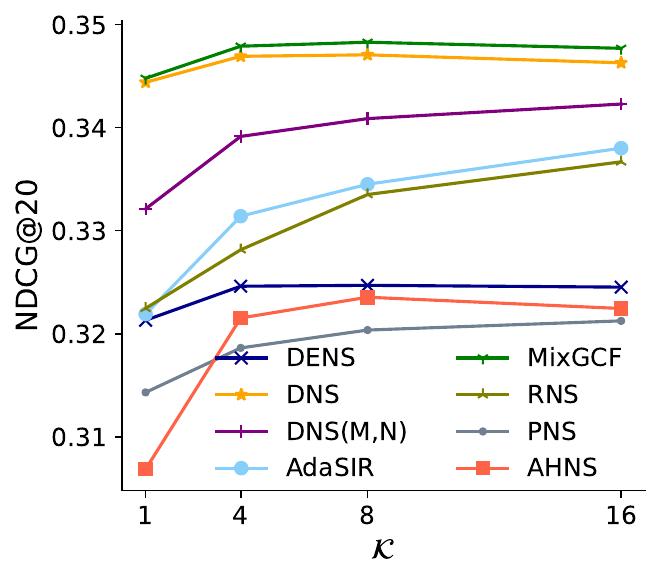}
        \caption{ML~1M -- LightGCN.}
        \label{fig:ml_light}
    \end{subfigure}
\hspace{0.4em}
    \begin{subfigure}{0.22\linewidth}
        \includegraphics[height=0.85\linewidth]{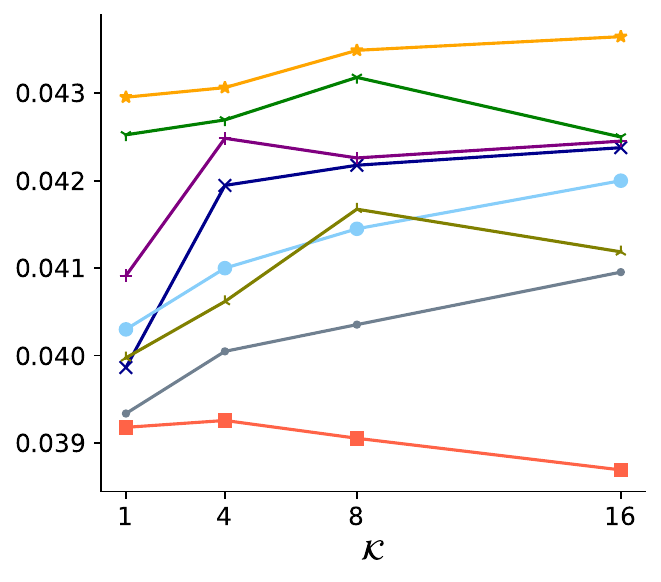}
        \caption{Amazon Beauty -- LightGCN.}
        \label{fig:beauty_light}
    \end{subfigure}
\hspace{0.4em}
    \begin{subfigure}{0.22\linewidth}
        \includegraphics[height=0.85\linewidth]{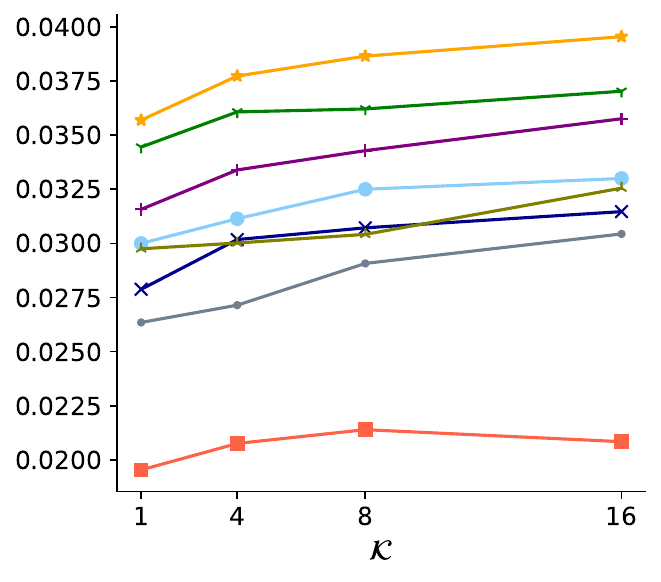}
        \caption{Book Crossing -- LightGCN.}
        \label{fig:book_light}
    \end{subfigure}
\hspace{0.4em}
    \begin{subfigure}{0.22\linewidth}
        \includegraphics[height=0.85\linewidth]{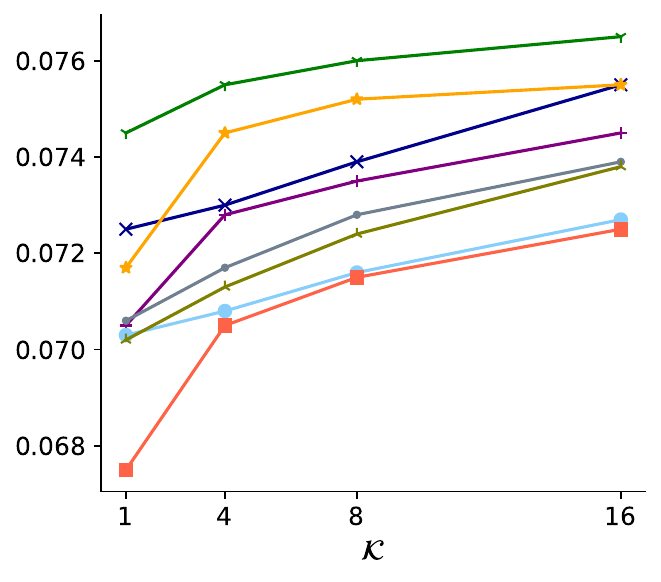}
        \caption{Yelp~2022 -- LightGCN.}
        \label{fig:pinterest_light}
    \end{subfigure}
    \caption{Performance of the recommenders with different negative sampling strategies under different negative sampling ratios across datasets. Samplers in Panels (\subref{fig:ml_mf}--\subref{fig:pinterest_mf}) are integrated with MF, and in Panels (\subref{fig:ml_light}--\subref{fig:pinterest_light}) with LightGCN.}
    \label{fig:increase_k_agg}
\end{figure*}

\subsection{Increasing Negative Ratio Improves Performance for Most Samplers (RQ2)}%

Next, we investigate whether negative samplers can benefit from increased negative sampling ratios. Following common practice~\cite{huang2021mixgcf,chen2023revisiting}, we set $\mathcal{K} \in \{1, 4, 8, 16\}$ and show the corresponding results for all the samplers in Figure~\ref{fig:increase_k_agg}. %
We observe that for most negative samplers, increasing the ratio enhances their average utility. %
Simple methods such as RNS enjoy a higher degree of improvement (generally more than 5\%), whereas the gains for samplers like DNS and MixGCF are relatively small (less than 5\%) and become marginal when increasing $\mathcal{K}$ from $8$ to $16$. Sometimes a large $\mathcal{K}$ can even harm the performance. For example, the performance of DNS and DNS(M,~N) in Figure~\ref{fig:ml_mf} decreases when $\mathcal{K}$ increases from $8$ to $16$. Similarly, MixGCF and DNS in Figure~\ref{fig:ml_light} get worse performance when changing $\mathcal{K}$ from $4$ to $8$ or $16$. We posit that when sampling more negative items, informative samples are more likely to be included and they yield bigger gradient updates. %
This is more beneficial for static samplers since they do not attempt to include informative negatives by design. But a bigger ratio is not always the best configuration as excessive negative items can introduce too much noise, especially in hard negative samplers.  

\begin{figure*}[t]
    \centering
    \begin{subfigure}{0.22\linewidth}
        \includegraphics[height=0.85\linewidth]{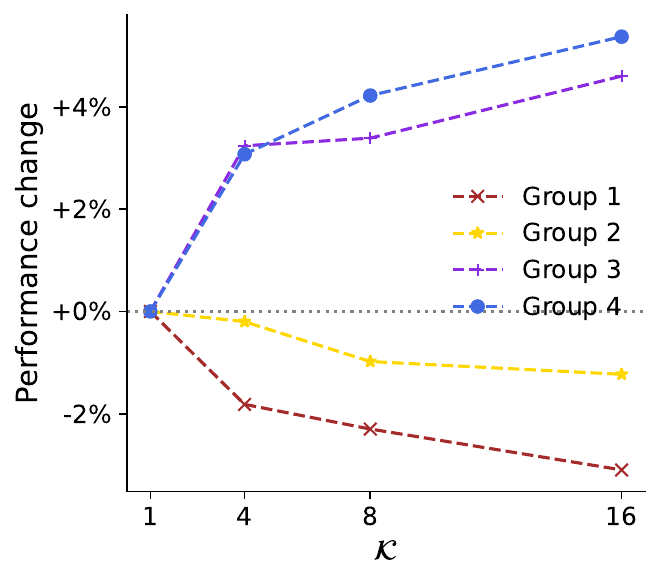}
        \caption{Amazon Beauty -- DNS.}
        \label{fig:beauty_dns_mf}
    \end{subfigure}
\hspace{0.4em}
    \begin{subfigure}{0.22\linewidth}
        \includegraphics[height=0.85\linewidth]{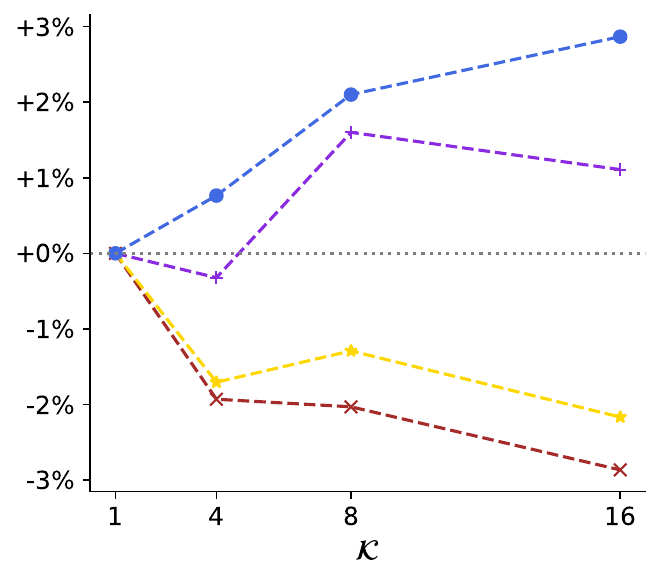}
        \caption{Amazon Beauty -- DNS(M,~N).}
        \label{fig:beauty_dnsmn_mf}
    \end{subfigure}
\hspace{0.4em}
    \begin{subfigure}{0.22\linewidth}
        \includegraphics[height=0.85\linewidth]{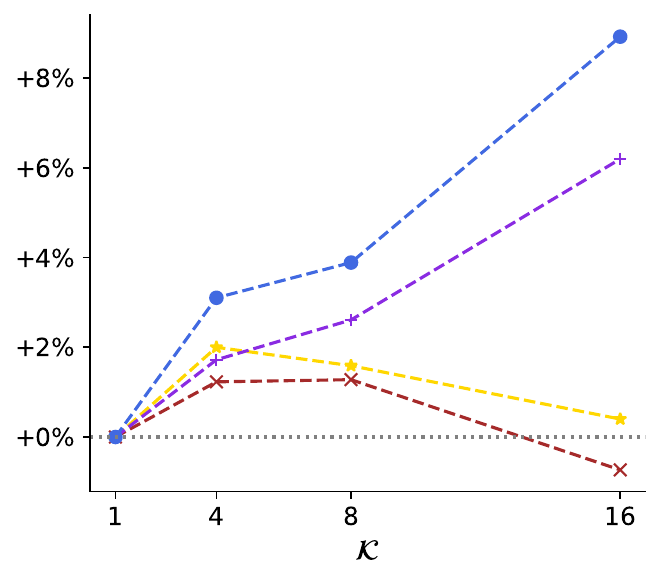}
        \caption{Amazon Beauty -- RNS.}
        \label{fig:beauty_rns_mf}
    \end{subfigure}
\hspace{0.4em}
    \begin{subfigure}{0.22\linewidth}
        \includegraphics[height=0.85\linewidth]{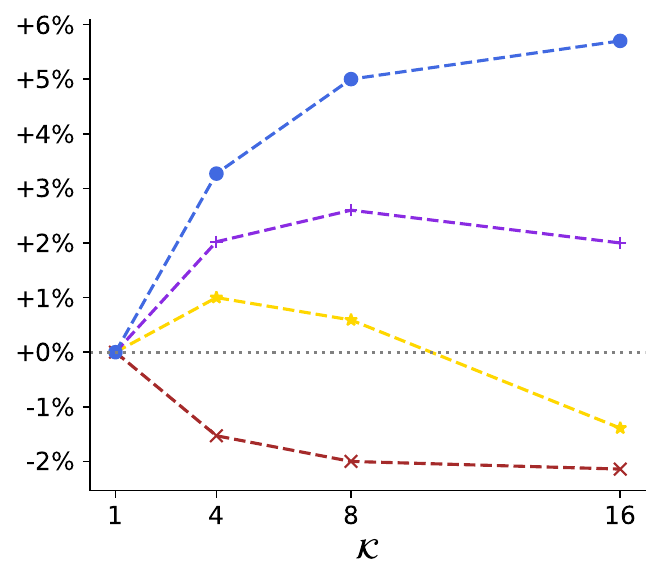}
        \caption{Amazon Beauty -- AdaSIR.}
        \label{fig:beauty_adasir_mf}
    \end{subfigure}
\\
    \begin{subfigure}{0.22\linewidth}
        \includegraphics[height=0.85\linewidth]{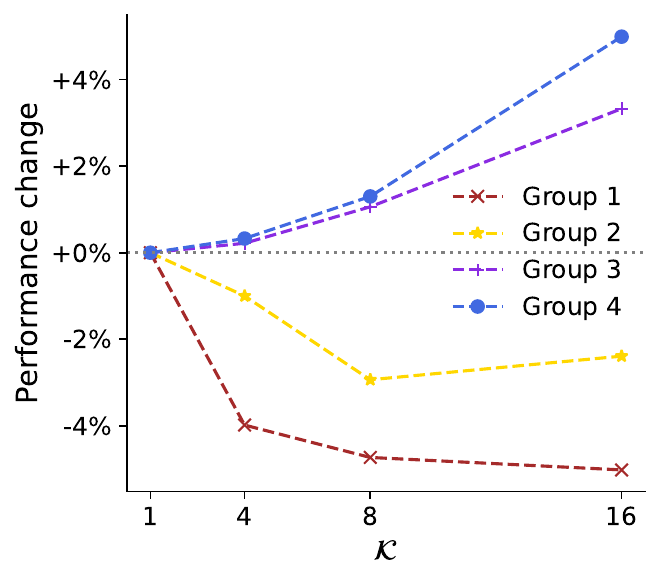}
        \caption{Book Crossing -- DNS.}
        \label{fig:book_dns_mf}
    \end{subfigure}
\hspace{0.4em}
    \begin{subfigure}{0.22\linewidth}
        \includegraphics[height=0.85\linewidth]{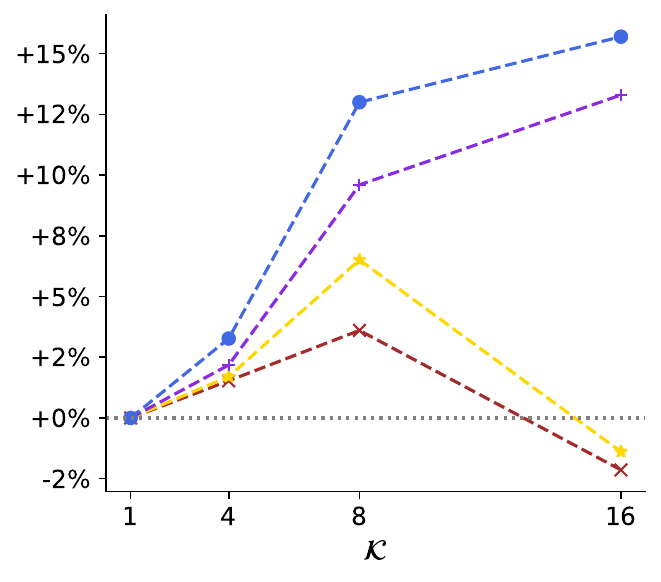}
        \caption{Book Crossing -- DNS(M,~N).}
        \label{fig:book_dnsmn_mf}
    \end{subfigure}
\hspace{0.4em}
    \begin{subfigure}{0.22\linewidth}
        \includegraphics[height=0.85\linewidth]{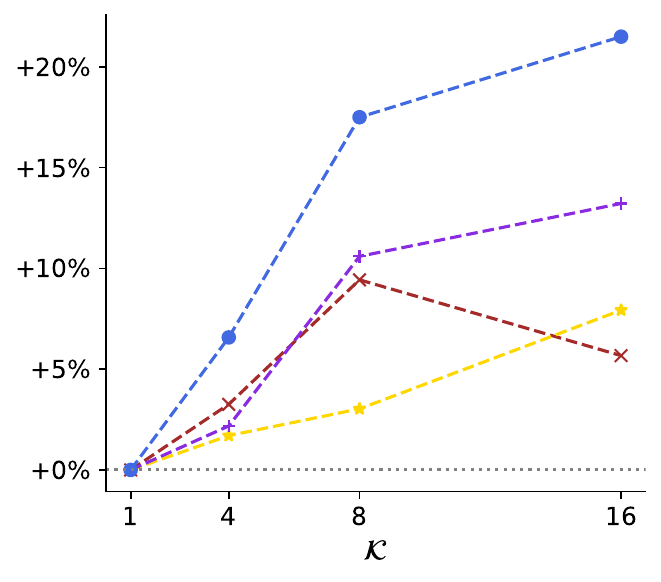}
        \caption{Book Crossing -- RNS.}
        \label{fig:book_rns_mf}
    \end{subfigure}
\hspace{0.4em}
    \begin{subfigure}{0.22\linewidth}
        \includegraphics[height=0.85\linewidth]{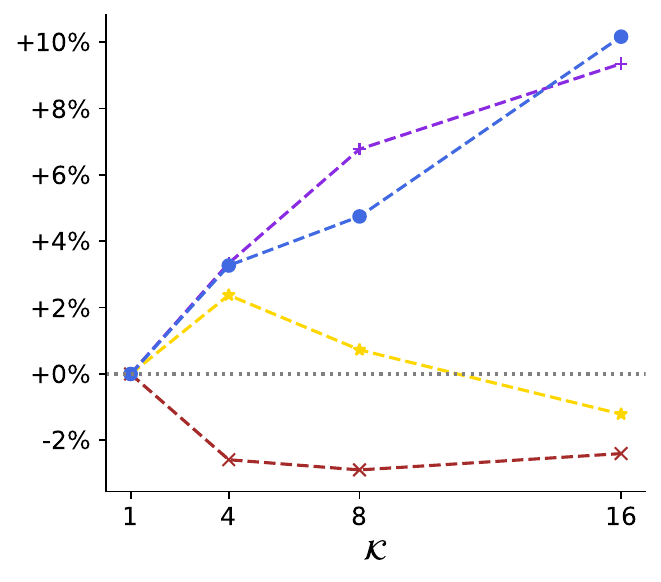}
        \caption{Book Crossing -- AdaSIR.}
        \label{fig:book_adasir_mf}
    \end{subfigure}
    \caption{Relative NDCG change for each user group %
    when $\mathcal{K}>1$ compared to performance at $\mathcal{K}=1$. For brevity, we report results for four representative samplers integrated with MF on two datasets. %
    Full results 
     -- which follow the same patterns -- are available in the dedicated code repository. %
    Group 1 captures the least and Group 4 the most active users.} %
    \label{fig:increase_k_group}
\end{figure*}

More notably, the relative performance ranking among samplers is not consistently maintained as evidenced by many crossing lines in Figure~\ref{fig:increase_k_agg}. For instance, in Figure~\ref{fig:pinterest_mf}, DENS and DNS(M,~N) surpass DNS as the best and second-best samplers when $\mathcal{K}$ increases. Similarly, DENS outperforms DNS as the best sampler in Figure~\ref{fig:book_mf} when $\mathcal{K}>1$. These observations highlight that benchmarking negative samplers exclusively for $\mathcal{K}=1$ is insufficient. %
Future research therefore ought to report performance of competitive samplers across various values of $\mathcal{K}$ to ensure a comprehensive and fair evaluation.  %

In our experiments, we limit $\mathcal{K}$ to $16$ as its larger values offer negligible performance gains while significantly increasing computation time per epoch despite requiring fewer epochs for convergence. %
The substantial rise in the total training time outweighs the marginal performance gains beyond $\mathcal{K}=16$. 
For example, with DNS and MF on ML~1M, an epoch takes 35 seconds at $\mathcal{K}=1$ and 406 seconds at $\mathcal{K}=16$. This aligns with prior work, which reported that  %
sampling more negative items increases the computation time~\cite{chen2017sampling}. 

In summary, increasing $\mathcal{K}$ improves the performance of negative samplers, however the relative performance ranking among samplers can fluctuate. %
In some cases a medium value of $\mathcal{K}$ is more effective as performance can degrade for larger values of $\mathcal{K}$. %

\subsection{Increasing Negative Ratio Further Disadvantages Inactive Users (RQ3)}%

After demonstrating that increasing $\mathcal{K}$ can improve the average performance of negative samplers,
we now examine if this improvement is equally distributed across sub-populations. 
Figure~\ref{fig:increase_k_group} shows the degree of performance change received by the four user groups when $\mathcal{K} > 1$ compared to $\mathcal{K}=1$. The results indicate an unequal distribution of performance gains across the groups. %

Specifically, Group~4 -- comprising the most active users -- receives a consistent increase in performance, with the best performance always achieved at $\mathcal{K}=16$. 
In most cases, its degree of gain surpasses that of the other groups, as shown by the blue line in Figures~\ref{fig:increase_k_group}\subref{fig:beauty_dnsmn_mf}--\subref{fig:book_rns_mf}, which remains above the lines of the other groups. %
Similarly, Group~3 -- the second most active group -- also receives performance boost, which often gets larger as $\mathcal{K}$ increases. %
This is captured by its line (purple) generally staying above the 0\% mark.  %
Notably, from RQ1 we know that Group~4 already receives above-average performance from all the negative samplers under $\mathcal{K}=1$. %
Combined with our findings for RQ3, which show that Group~4 continuously receives more gains as $\mathcal{K}$ increases, we observe that increasing $\mathcal{K}$ helps ``the rich get richer''. %

While active users benefit from this setup, inactive users do not always experience performance gains and sometimes suffer performance degradation. For example, in Figures~\ref{fig:beauty_dns_mf}, \subref{fig:beauty_dnsmn_mf} and~\subref{fig:book_dns_mf}, performance for Groups~1 and 2 under $\mathcal{K}>1$ is always worse than when $\mathcal{K}=1$, suggesting that bigger ratios disadvantage inactive users. %
Even when there is a performance gain for Group~1, the benefit peaks at a medium-valued $\mathcal{K}$, such as $\mathcal{K}=4$ in Figure~\ref{fig:beauty_rns_mf} and $\mathcal{K}=8$ in Figure~\ref{fig:book_dnsmn_mf}, and declines when $\mathcal{K}$ gets too large. %

\begin{table}[b]
\centering
\footnotesize
    \setlength{\tabcolsep}{3pt}%
\caption{Kendall's $W$ under varying $n$ and partition criteria. %
For brevity, we only report samplers from Figure~\ref{fig:increase_k_group} with MF; %
full results with similar patterns are in the code repository. %
}
    \begin{tabular}{@{}p{0.5cm}lccccrc@{}}
    \toprule
     &  & \multicolumn{4}{c}{Number of Groups ($n$)} & \multicolumn{2}{c}{Alternative Criteria} \\
     \cmidrule(lr){3-6} \cmidrule(lr){7-8}
     & Sampler & $n=2$ & $n=4$ & $n=7$ & $n=10$ & \multicolumn{1}{c}{\hspace{0.4cm}ATP} & IF \\ 
     \midrule
    \multirow{4}{*}{\rotatebox[origin=c]{90}{\parbox[c]{0.8cm}{\centering Amazon Beauty}}}
         & DNS      & 1.00 & 0.91 & 0.87 & 0.86 & 0.05 & 0.04 \\
         & DNS(M,~N) & 1.00 & 1.00 & 0.94 & 0.98 & 0.11 & 0.13 \\
         & RNS      & 1.00 & 0.91 & 0.89 & 0.88 & 0.06 & 0.10 \\
         & AdaSIR   & 1.00 & 1.00 & 0.93 & 0.90 & 0.15 & 0.04 \\
    \midrule
     \multirow{4}{*}{\rotatebox[origin=c]{90}{\parbox[c]{0.8cm}{\centering ML~1M}}} 
         & DNS      & 1.00 & 1.00 & 0.91 & 0.93 & 0.08 & 0.07 \\
         & DNS(M,~N) & 1.00 & 1.00 & 0.95 & 0.98 & 0.10 & 0.02 \\
         & RNS      & 1.00 & 0.87 & 0.88 & 0.82 & 0.03 & 0.05 \\
         & AdaSIR   & 1.00 & 0.91 & 0.90 & 0.92 & 0.04 & 0.05 \\
    \bottomrule
    \end{tabular}
    \label{tab:ablation_table}
\end{table}

The persistence of unequal distribution is also evident in Table~\ref{tab:ablation_table}, where Kendall's $W$ is $1$ or close to $1$ for $n=4$. This shows that the line representing Group~4 is consistently above the lines of the other groups; %
the line for Group~3 is above those of Groups~1 and 2; and Group~2 is above Group~1. %
Despite this, such disparity is often hidden in mean utility metrics, where overall gains mask the performance decline for under-represented groups.
For instance, while DNS's average performance in Figure~\ref{fig:beauty_mf} improves with increasing $\mathcal{K}$, the upward trend is primarily driven by Groups~3 and 4 (Figure~\ref{fig:beauty_dns_mf}), whereas Groups~1 and 2 suffer a decline when $\mathcal{K}>1$.
Our findings echo the call for pessimistic evaluation of recommenders, focusing on worst-case utility to promote social good~\cite{diaz2024pessimistic}.

In summary, improving overall performance when $\mathcal{K}>1$ does not guarantee that all users benefit equally. In fact, inactive users often receive negligible performance gains or even suffer losses, while active users see substantial improvements.  
This disparity also indicates that a uniform negative sampling ratio for all groups may not be the optimal setup. Therefore, in Section~\ref{sec:rq4} we investigate %
assigning different ratios to distinct user groups (RQ4).  %

\subsection{Ablation Study}\label{sec:ablation}%

There are two factors affecting the robustness of our findings: the number of user groups $n$ and the user partition strategy. %
Here, we demonstrate that our findings remain consistent regardless of $n$, and validate user activity level as a suitable partitioning criterion.

\paragraph{Number of Groups}%
Since RQ1 and RQ3 focus on group-level performance, $n$ becomes a key factor. %
We test $n\in\{2,4,7,10\}$ while keeping the same user partitioning criterion. For RQ1, we find that when $n\ge4$, only the top active 20--30\% of users receives above-average performance; the remaining 70--80\% gets below-average performance. %
When $n=2$, the top 50\% of users (by activity level) receives significantly higher utility than the bottom 50\%. For RQ3, Table~\ref{tab:ablation_table} shows that Kendall's $W$ is consistently $1$ or close to $1$ across all the values of $n$. %
This indicates that more active users always receive a higher performance gain as explained in Definition~\ref{def:ranking}. Consequently, RQ1 and RQ3 still hold regardless of the value of $n$. %

\paragraph{Alternative User Partitioning Criteria}
Next, %
we group users based on two alternative criteria: their Active Time Period (ATP) and Interaction Frequency (IF). %
ATP is the number of days between a user's first and last interaction, while $\text{IF}=\text{user activity level}/\text{ATP}$. Since calculating these two criteria requires interaction timestamps, we are only able to compute them for Amazon Beauty and ML~1M. %
Table~\ref{tab:ablation_table} shows that when using ATP or IF, $W$ is consistently close to $0$, suggesting that there is no difference in the disparity of performance gain when $\mathcal{K}$ increases. 
This indicates that ATP and IF are not useful group partitioning criteria, %
hence our findings are %
strongly linked to the nature of group membership.  %
This result is expected as none of the recommenders or samplers we evaluated are temporal or leverage time-based features. 
Future research could thus explore how temporal factors influence performance disparity under negative sampling in sequential recommenders. %

\subsection{Theoretical Analysis}%
We now present a theoretical analysis of our key findings in RQ3 based on the framework introduced by \citet{wu2021rethinking} that quantifies training effectiveness for a given $\mathcal{K}$. Their method calculates the proportion of informative (``good'') and noisy (``bad'') negative samples among the sampled items, and computes the resulting training effectiveness. 
While their work focused on dataset-level effectiveness, we extend this framework to measure training effectiveness for each user group in Amazon Beauty as a representative example.

\begin{figure}[t]
    \begin{subfigure}{0.32\columnwidth}
        \includegraphics[height=.8\linewidth]{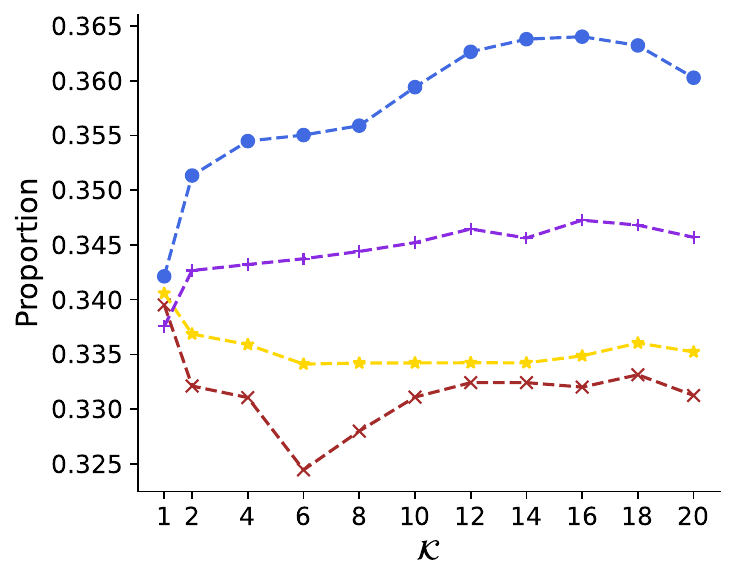}
        \caption{Good samples.}
        \label{fig:beauty_good}
    \end{subfigure}
    \begin{subfigure}{0.32\columnwidth}
        \includegraphics[height=.8\linewidth]{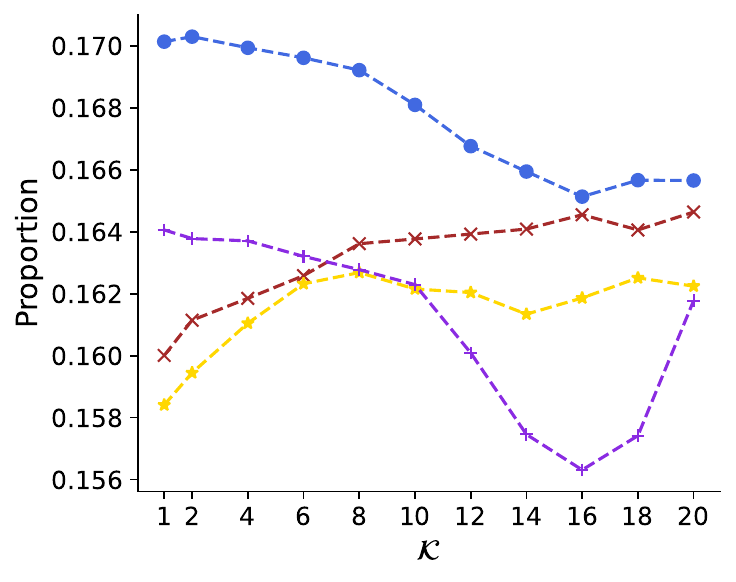}
        \caption{Bad samples.}
        \label{fig:beauty_bad}
    \end{subfigure}
    \begin{subfigure}{0.32\columnwidth}
        \includegraphics[height=.8\linewidth]{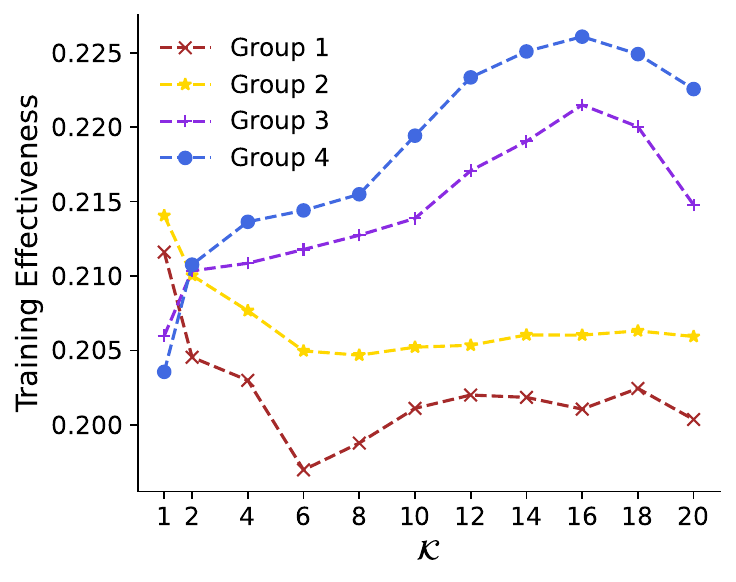}
        \caption{Effectiveness.}
        \label{fig:beauty_effectiveness}
    \end{subfigure}
    \caption{
    Proportion of (\subref{fig:beauty_good}) good and (\subref{fig:beauty_bad}) bad samples as well as (\subref{fig:beauty_effectiveness}) training effectiveness for each user group in the Amazon Beauty dataset with DNS and MF. Calculation details can be found in the original paper~\cite{wu2021rethinking}.
    }
    \label{fig:beauty_effectiveness_main}
\end{figure}

As Figure~\ref{fig:beauty_effectiveness_main} shows, when $\mathcal{K}$ increases, the negative items sampled for active user groups (Groups~3 and 4) contain more good and fewer bad samples, which in turn improves their training effectiveness. %
However, when $\mathcal{K}>16$, the effectiveness starts to decline due to an increase in bad samples, which harms the training process. In contrast, less active users (Groups~1 and 2) see a consistent drop in training effectiveness as soon as $\mathcal{K}>1$, with fewer good and more bad samples included in their negative samples. 

These trends in theoretical training effectiveness align with the empirical performance changes shown in Figure~\ref{fig:beauty_dns_mf}, where performance increases for Groups~3 and 4 and decreases for Groups~1 and 2. 
This alignment also confirms that user activity level is an appropriate grouping criterion as it captures meaningful differences in both training effectiveness and performance changes across groups.

The observed variation in training effectiveness further explains why smaller $\mathcal{K}$ values are more suitable for inactive user groups, while larger $\mathcal{K}$ values benefit active user groups. 
Namely, since limited data are available for inactive users, it is easier to learn a precise decision boundary for them with only a small number of negative samples. %
Using an excessive number of negative samples can introduce too much noise to their data representation, decreasing the performance. 
Conversely, active users have more data, %
so a larger $\mathcal{K}$ is needed to exploit sufficient information and learn a good decision boundary. 
Their data representation is learnt from abundant data, thus it is more robust to higher levels of noise introduced by negative samples. %
Similarly, when $\mathcal{K}$ becomes excessively large, the larger number of noisy samples starts to harm the training. %

\section{Optimising Group-specific Ratio (RQ4)}\label{sec:rq4}%
Building upon our findings for the first three research questions %
we now explore RQ4, assessing whether using different negative ratios for different user groups improves the performance compared to a uniform ratio; and  %
if so, how to determine the optimal ratio for each group. 
Selecting $\mathcal{K}$ for each group poses a combinational optimisation challenge. The number of possible combinations grows exponentially as the possible values of $\mathcal{K}$ and the number of groups increase. Evaluating each combination also requires training a recommender from scratch, making grid search computationally infeasible.
We therefore treat each group's $\mathcal{K}$ as a hyperparameter and use a hyperparameter optimisation method to identify the optimal $\mathcal{K}$ for each user group. By leveraging our earlier findings, we are able to accelerate this optimisation process. Our results show that group-specific ratios improve recommender performance across all the negative samplers. %

\begin{table*}[t]
\footnotesize
\setlength\tabcolsep{3pt}
\caption{Recommendation utility of the eight negative samplers paired with the two recommenders assessed using NDCG@20 and Recall@20; the table also reports Recall@20 for the least active group (Recall@20-min). %
The metrics are calculated for a global ratio with $\mathcal{K}=16$ and a group-specific ratio optimised using informed Hyperband (the HB column).} %
\begin{tabular}{@{}p{0.25cm}lr>{\bfseries}rr>{\bfseries}rr>{\bfseries}rr>{\bfseries}rr>{\bfseries}rr>{\bfseries}rr>{\bfseries}rr>{\bfseries}rr>{\bfseries}r@{}}
\toprule
 &  & \multicolumn{6}{c}{ML~1M} & \multicolumn{6}{c}{Amazon Beauty} & \multicolumn{6}{c}{Book Crossing} \\
 \cmidrule(lr){3-8}\cmidrule(lr){9-14} \cmidrule(lr){15-20}
 &  & \multicolumn{2}{c}{NDCG@20} & \multicolumn{2}{c}{Recall@20} & \multicolumn{2}{c}{Recall@20-min} & \multicolumn{2}{c}{NDCG@20} & \multicolumn{2}{c}{Recall@20} & \multicolumn{2}{c}{Recall@20-min} & \multicolumn{2}{c}{NDCG@20} & \multicolumn{2}{c}{Recall@20} & \multicolumn{2}{c}{Recall@20-min}\\
 \cmidrule(lr){3-4} \cmidrule(lr){5-6} \cmidrule(lr){7-8}
 \cmidrule(lr){9-10} \cmidrule(lr){11-12} \cmidrule(lr){13-14}
 \cmidrule(lr){15-16} \cmidrule(lr){17-18} \cmidrule(lr){19-20}
 & Sampler & $\mathcal{K}$=16 & \normalfont{HB} & $\mathcal{K}$=16 & \normalfont{HB} & $\mathcal{K}$=16 & \normalfont{HB} & $\mathcal{K}$=16 & \normalfont{HB} & $\mathcal{K}$=16 & \normalfont{HB} & $\mathcal{K}$=16 & \normalfont{HB} & $\mathcal{K}$=16 & \normalfont{HB} & $\mathcal{K}$=16 & \normalfont{HB} & $\mathcal{K}$=16 & \normalfont{HB} 
 \\ \midrule
 \multirow{7}{*}{\rotatebox[origin=c]{90}{\parbox[c]{2cm}{\centering MF}}} 
 & RNS & 0.322 & 0.330 & 0.225 & 0.229 & 0.121 & 0.137 & 0.032 & 0.035 & 0.059 & 0.063 & 0.043 & 0.048 & 0.028 & 0.031 & 0.045 & 0.048 & 0.038 & 0.040 \\
 & PNS & 0.322 & 0.328 & 0.224 & 0.227 & 0.121 & 0.128 & 0.032 & 0.038 & 0.058 & 0.061 & 0.041 & 0.046 & 0.029 & 0.035 & 0.037 & 0.041 & 0.035 & 0.041 \\
 & DNS & 0.337 & 0.341 & 0.235 & 0.238 & 0.125 & 0.131 & 0.037 & 0.040 & 0.071 & 0.075 & 0.055 & 0.061 & 0.038 & 0.041 & 0.059 & 0.062 & 0.052 & 0.058 \\
 & AdaSIR & 0.325 & 0.330 & 0.228 & 0.231 & 0.118 & 0.125 & 0.035 & 0.037 & 0.067 & 0.069 & 0.053 & 0.059 & 0.031 & 0.035 & 0.057 & 0.058 & 0.051 & 0.055 \\
 & DNS(M,~N) & 0.330 & 0.334 & 0.231 & 0.235 & 0.123 & 0.127 & 0.034 & 0.037 & 0.064 & 0.068 & 0.048 & 0.050 & 0.034 & 0.037 & 0.055 & 0.059 & 0.049 & 0.052 \\ 
 & DENS & 0.331 & 0.335 & 0.229 & 0.231 & 0.123 & 0.128 & 0.036 & 0.038 & 0.069 & 0.073 & 0.055 & 0.058 & 0.039 & 0.042 & 0.061 & 0.065 & 0.056 & 0.069 \\
 & AHNS & 0.296 & 0.312 & 0.215 & 0.252 & 0.120 & 0.158 & 0.034 & 0.037 & 0.062 & 0.066 & 0.051 & 0.057 & 0.025 & 0.031 & 0.037 & 0.045 & 0.035 & 0.049 \\
 \midrule
\multirow{8}{*}{\rotatebox[origin=c]{90}{\parbox[c]{2.0cm}{\centering LightGCN}}} 
 & RNS & 0.336 & 0.341 & 0.238 & 0.241 & 0.123 & 0.127 & 0.041 & 0.043 & 0.081 & 0.085 & 0.067 & 0.070 & 0.032 & 0.034 & 0.049 & 0.050 & 0.045 & 0.048\\
 & PNS & 0.321 & 0.331 & 0.232 & 0.237 & 0.122 & 0.130 & 0.041 & 0.043 & 0.080 & 0.082 & 0.064 & 0.068 & 0.030 & 0.032 & 0.049 & 0.051 & 0.041 & 0.045 \\
 & DNS & 0.346 & 0.348 & 0.243 & 0.246 & 0.127 & 0.131 & 0.044 & 0.045 & 0.085 & 0.086 & 0.070 & 0.076 & 0.039 & 0.041 & 0.058 & 0.060 & 0.049 & 0.054 \\
 & AdaSIR & 0.338 & 0.342 & 0.240 & 0.243 & 0.125 & 0.131 & 0.042 & 0.044 & 0.082 & 0.083 & 0.065 & 0.070 & 0.033 & 0.037 & 0.051 & 0.053 & 0.047 & 0.049 \\
 & DNS(M,~N) & 0.342 & 0.348 & 0.242 & 0.249 & 0.125 & 0.134 & 0.042 & 0.044 & 0.083 & 0.085 & 0.067 & 0.071 & 0.036 & 0.038 & 0.054 & 0.056 & 0.048 & 0.051 \\ 
 & MixGCF & 0.347 & 0.349 & 0.244 & 0.245 & 0.128 & 0.132 & 0.043 & 0.044 & 0.083 & 0.085 & 0.071 & 0.075 & 0.037 & 0.039 & 0.052 & 0.524 & 0.038 & 0.041 \\ 
 & DENS & 0.325 & 0.332 & 0.241 & 0.249 & 0.121 & 0.129 & 0.042 & 0.043 & 0.083 & 0.084 & 0.067 & 0.069 & 0.031 & 0.034 & 0.044 & 0.046 & 0.033 & 0.041 \\
 & AHNS & 0.324 & 0.339 & 0.239 & 0.241 & 0.114 & 0.165 & 0.038 & 0.040 & 0.078 & 0.080 & 0.069 & 0.071 & 0.021 & 0.028 & 0.029 & 0.033 & 0.028 & 0.036 \\
\bottomrule
\end{tabular}
\label{tab:results}
\end{table*}

\subsection{Hyperband with Informed Priors}
Current literature often uses a global $\mathcal{K}$ setting for all users without any justification. To improve upon this practice, we optimise group-specific ratios using Hyperband~\cite{li2018hyperband} -- a hyperparameter optimisation algorithm that finds suitable hyperparameters through a pure-exploration infinite-armed bandit. It adaptively allocates resources among randomly sampled hyperparameter configurations 
and discards poorly performing ones via an early-stopping strategy. 
We adapt Hyperband to evaluate combinations of different $\mathcal{K}$ values for different user groups and identify the optimal group-specific ratios that yield the best performance. 
We further enhance it by incorporating intelligent priors derived from RQ3. Specifically, we use smaller $\mathcal{K}$ values (e.g., $\mathcal{K}=4$) for inactive groups and larger values (e.g., $\mathcal{K}=12$) for more active groups as the priors of their ratios. To generate a candidate configuration, we sample $\mathcal{K}$ for every group from their corresponding prior distribution. %
The process is repeated to build a configuration pool for Hyperband to explore.

To ensure a consistent and fair evaluation, we constrain each group's $\mathcal{K}$ to a fixed range of integers between $1$ and $16$. 
This range is chosen to maintain computational feasibility; it also aligns with our previous finding that $\mathcal{K}>16$ harms training.
We compare our approach against a baseline where a uniform global ratio of $\mathcal{K}=16$ is applied across all user groups. $\mathcal{K}=16$ is chosen because it yields the best overall performance for most samplers.
We fix $n=4$ for consistency, but later show that our method improves performance regardless of the chosen $n$. %
Lastly, we run Hyperband without priors to highlight the benefits of using priors in the optimisation process. %

\subsection{Experimental Results}\label{sec:group-ratio-result}

We apply Hyperband with informed priors (InformedHB) to each negative sampler to find the best group-specific ratios. We compare the results with the global-ratio setup and report them in Table~\ref{tab:results}. %
In addition to NDCG@20 and Recall@20, we also report \mbox{Recall@20-min}, which measures the performance for the most disadvantaged group (Group~1). %
Recall@20-min was introduced in prior work as the worst-case analysis metric, offering insights into equal access to high-quality recommendations~\cite{diaz2024pessimistic,xu2024fairly};
unlike Recall@$\cdot$, which is aggregated over an entire population, this metric is aggregated across a sub-population. %
Each experiment is repeated three times and the average performance is reported. %
The exact group-specific ratios found by the optimiser for each method are provided in our code repository. Additional results for %
the Yelp~2022 dataset -- which follow similar patterns -- are included there as well. %

Specifically, %
we find that, for every negative sampler, using group-specific negative ratios consistently improves the performance %
by a large margin over a global ratio, confirming that it is a better approach for sampling-based recommenders. 
The relative improvement in NDCG and Recall typically ranges from 2 to 10\% given that the performance of most samplers does not benefit significantly from larger $\mathcal{K}$ in any setup (refer back to Figure~\ref{fig:increase_k_agg}).
Our results also show that group-specific ratios are highly effective across all evaluated samplers, indicating that our hyperparameter optimisation strategy is broadly compatible with various negative samplers. %

Notably, the performance gain for the least active users is more significant as visible in Recall@20-min increasing more than Recall@20. %
For example, in Amazon Beauty, Recall@20-min increased by roughly 6\% while Recall@20 only by 3\% for most samplers.  %
We attribute this to inactive users being disproportionately affected under a high global negative sampling ratio. After tailoring the ratio to each user group, our method delivers greater gains to inactive users, thus promoting fairness without sacrificing overall utility. %

\paragraph{Empirical Complexity}
A key contribution of our work is the integration of priors into Hyperband to accelerate search convergence. %
Across all the datasets, we find that Hyperband without priors requires exploring up to nine times more configurations to find the same optimal setup as the informed version. %
Both informed and uninformed Hyperband converge to the same optimal configuration, demonstrating that using priors ensures robust results while significantly reducing computational costs.

\paragraph{Different Number of Groups}%

To verify the generalisability of our method across different number of sub-populations, we run informed Hyperband with users split into $n=2$ and $n=7$ groups. For brevity, we include the results in our code repository.  %
We find that a more refined user grouping allows a more diverse set of ratios and can improve performance. However, when the number of groups becomes excessive (e.g., $n>4$), the variation between groups diminishes and similar ratios are assigned to multiple groups, with the improvement becoming marginal. 
Furthermore, increasing $n$ beyond $10$ significantly slows down the convergence of Hyperband, reducing its practicality. 
Thus, in the absence of background knowledge on the optimal number of user groups, setting $n=4$ offers a practical balance between performance and efficiency. %

\paragraph{Insights for Cold-start Problem}%

In our user grouping, inactive users resemble ``cold-start'' users with limited interactions, while active users are ``warm'' users with more engagement. Although the evaluated samplers are not designed for cold-start scenarios, RQ4 offers valuable insights for this setting; namely, to assign fewer negative samples per positive interaction to less active users, and gradually increase the ratio as their activity grows. Future work could thus extend our evaluation to cold-start samplers. %

\section{Conclusion}

Providing equally satisfactory recommendations for all user groups is key to trustworthy recommenders. In this work, we systematically evaluated eight negative sampling methods for recommenders from a user-centred perspective. We found that inactive users, who constitute the majority of many user-bases, receive unsatisfactory results, whereas active users, who often are the minority, get superior performance across all the samplers.  
Increasing the negative sampling ratio can enhance the overall performance, but it disproportionately benefits active users while degrading the experience of inactive ones. 
We proposed a group-specific negative sampling ratio that assigns smaller ratios to inactive user groups and larger ratios to active groups. Extensive experiments showed that the group-specific ratio improves the sampler performance over the global ratio, with particularly significant gains for inactive users.

\section*{Acknowledgements}
This research was conducted by the ARC Centre of Excellence for Automated Decision-Making and Society (Project No.\ CE200100005), funded by the Australian Government through the Australian Research Council. Additional support was provided by the Hasler Foundation (Grant No.\ 23082).

\bibliographystyle{ACM-Reference-Format}
\bibliography{reference}

\end{document}